\documentclass[english,showpacs,twocolumn,aps,reprint]{revtex4-1}

\usepackage{mathptmx}
\usepackage[normalem]{ulem}
\usepackage{xcolor}

\usepackage[T1]{fontenc}
\usepackage[latin9]{inputenc}
\usepackage{color}
\usepackage{amsmath}
\usepackage{amssymb}
\usepackage{hyperref}
\usepackage{graphics}
\usepackage{graphicx}
\usepackage{epsfig}
\usepackage{bm}
\usepackage{epstopdf}
\usepackage{mathtools}
\usepackage{soul}
\usepackage{siunitx}
\usepackage{subfigure}   
\usepackage{xmpmulti}

\usepackage{babel}
\begin{document}

\title{
{\color{black}Transition from Zonal Flows to Streamer like structures and associated edge Fluctuations}}

\author{Tanmay Karmakar\textsuperscript{1,2}}
\email{tanmay.karmakar@ipr.res.in}

\author{Rosh Roy\textsuperscript{1,2}}

\author{Lavkesh Lachhvani\textsuperscript{1,2}}
\author{Raju Daniel\textsuperscript{1,2}}
\author{Bhoomi Khodiyar\textsuperscript{1,2}} 
\author{Prabal K. Chattopadhyay\textsuperscript{1,2}}
\author{Abhijit Sen\textsuperscript{1,2}}
\author{Sayak Bose\textsuperscript{3,a}}

\affiliation{\textsuperscript{1}Institute for Plasma Research, HBNI, Bhat, Gandhinagar, 382428, India} 

\affiliation{\textsuperscript{2}Homi Bhabha National Institute, Anushaktinagar, Mumbai, Maharastra 400094, India}

\affiliation{\textsuperscript{3}Columbia Astrophysics Laboratory, Columbia University, 550 West 120th Street, New York, New York 10027, USA}

\affiliation{\textsuperscript{a}Present address: Princeton Plasma Physics Laboratory, Princeton, New Jersey 08540, USA.}
 
\begin{abstract}  
\noindent
We report experimental observations of a controlled transition from a zonal-flow (ZF) dominated regime to a coexistence regime of ZFs and streamers, and finally to a streamer-dominated state in a linear magnetized plasma column. The controlling parameter is the ion-neutral collision frequency. 
At low collisionality ($2\times10^{-5}~\mathrm{mbar}$), the plasma turbulence is dominated by coherent ZFs (600--700~Hz) that are nonlinearly driven by drift-wave fluctuations. With increasing collisionality ($5\times10^{-4}~\mathrm{mbar}$), the ZF growth is reduced and streamers emerge through a nonlinear coupling of neighboring drift modes mediated by a mediator mode. At high collisionality ($2\times10^{-3}~\mathrm{mbar}$), ZFs are strongly damped and the turbulence becomes streamer-dominated. For each of these turbulent states the corresponding edge fluctuations transition from coherent, symmetric to intermittent, asymmetric fluctuations with enhanced low-frequency content and larger spatial scales that can result in convective transport. Our results demonstrate the possibility of selective excitation of ZFs and streamers by regulating their  collisional damping and showcase  the ion--neutral collision frequency as an effective control knob for regulating turbulent structures and edge transport in magnetized plasmas.

\end{abstract}
\maketitle

\begin{figure*}[!hbt]
\includegraphics[width=\linewidth]{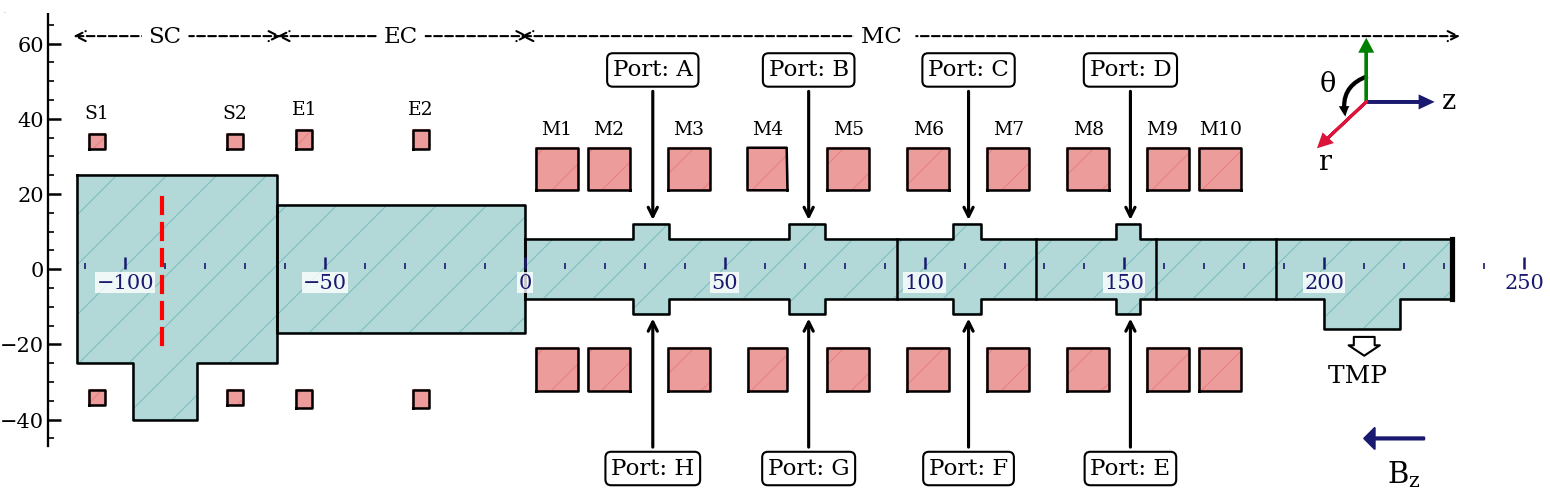} 
\caption{Schematic of the IMPED device with various radial ports.}    
\label{IMPED schematics}  
\end{figure*}

\begin{figure}[!hbt]
\includegraphics[width=\linewidth]{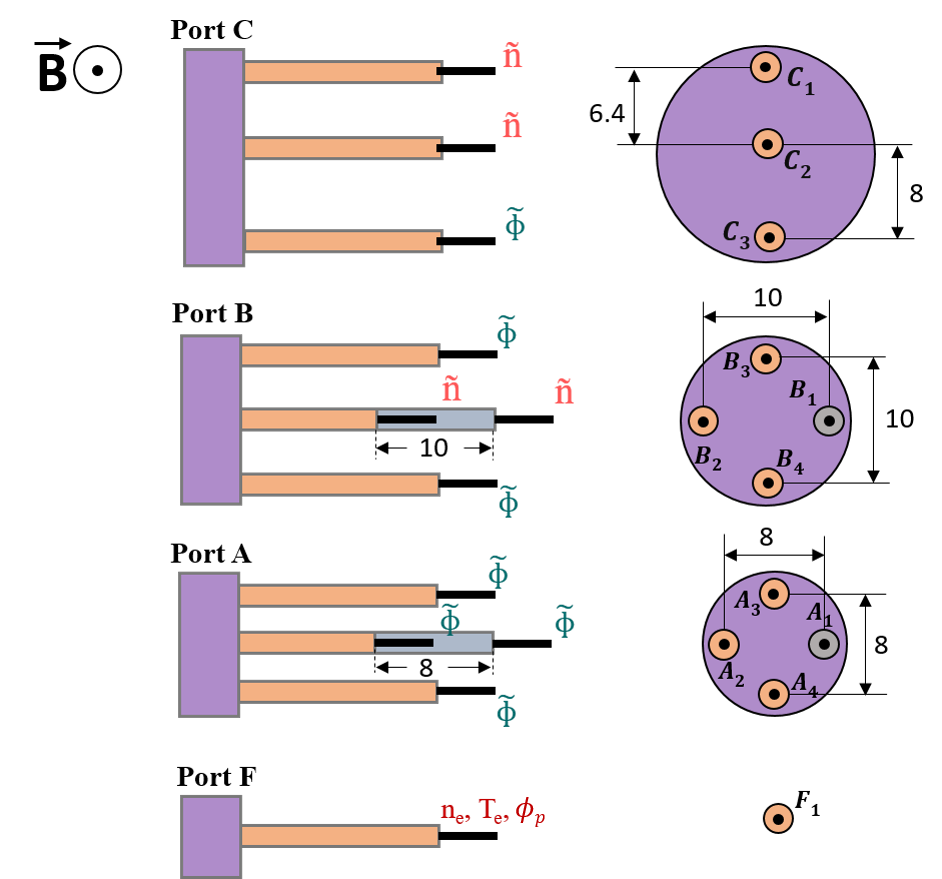} 
\caption{Probe configuration for measuring equilibrium profiles and fluctuation spectra. 
 }    
\label{Probe diagnostics}  
\end{figure}
  
\section{Introduction}
\noindent 
Edge turbulence in magnetized plasmas is intrinsically nonlinear and capable of self-organizing into meso-scale coherent structures that critically regulate cross-field transport \cite{horton1999drift, holland_2006observation}. In systems possessing equilibrium density and temperature gradients, drift-wave turbulence develops at micro-scales of the order of the ion sound Larmor radius $\rho_s$, leading to anomalous radial particle and momentum flux. However, nonlinear interactions among drift modes redistribute fluctuation energy across scales, giving rise to larger coherent structures that can either suppress or enhance transport \cite{kasuya2008selective}. Among these, zonal flows (ZFs) and streamers represent two fundamentally distinct states of turbulence self-organization. Zonal flows, characterized by $k_\theta = 0$ and finite $k_r$, were first predicted theoretically through modulational instability\cite{fujisawa_2004identification,tynan_2006observation} and inverse spectral energy cascade mechanisms, where Reynolds stress transfers energy from broadband drift-wave turbulence to azimuthally symmetric shear flows. These low-frequency or nearly stationary $E \times B$ shear flows act back on the turbulence by decorrelating eddies, reducing fluctuation amplitudes, and suppressing radial transport \cite{Terry_2000suppression,diamond_2004review, itoh2006physics, diamond_mean_flow_1991theory}. Subsequent gyrokinetic simulations \cite{kobayashipreypredetor2015direct, morel2014characterization} and analytical models \cite{diamond2011vorticity}  established the predator prey-type interaction between drift waves and ZFs, highlighting their central role in turbulence regulation. Experimentally, zonal flows have been identified in various linear and toroidal plasma devices through observations of low-frequency potential oscillations, radially sheared poloidal flows, and enhanced cross-coherence between floating potential and poloidal velocity fluctuations, confirming their shear-regulating character at the plasma edge \cite{fujisawa_2004identification, fujisawa_2008review, mckee_2003experimental, sokolov_2006observation, shen_2016observations, xu_2009blob}. In contrast, streamers were predicted in theoretical and numerical studies as anisotropic structures emerging from nonlinear mode coupling when spectral energy transfer favors elongation in the radial direction \cite{kasuya_streamers_2010selective, kasuya2008selective}. Unlike ZFs, streamers are characterized by $k_r \approx 0$ and finite $k_\theta$, forming radially extended convective cells \cite{yamada2010observation} that enhance radial transport. Simulations of drift-wave turbulence demonstrated that when inverse cascade toward $k_\theta=0$ is weakened or suppressed, energy can accumulate in modes with small radial wavenumber, leading to streamer formation. Experimentally, streamer-like structures have been reported \cite{kobayashi2017phenomenological, yamada_2008anatomy} as quasi-periodic bunching of drift waves forming radially elongated structures. These structures are long-lived compared to the turbulence decorrelation time and are known to produce enhanced particle flux and intermittent burst events in edge plasmas \cite{kin2019observations}. These observations established streamers as transport-enhancing meso-scale structures, fundamentally different from the shear-regulating nature of ZFs \cite{fujisawa_2008review}. 
Streamers have also been observed in tokamak plasmas. In the JIPP T-IIU, streamer eddies were detected in the outer plasma region, where the waveforms appear as pulses of complex shape with sharp fronts \cite{hamada2006streamers}. These features are consistent with streamer simulations reported by Garbet \textit{et al.}, showing similar steepened pulse structures. Ghizzo \textit{et al.} \cite{ghizzo2010streamer} showed that the interaction of streamer-like structures with plasma turbulence can drive an inverse cascade through nonlinear wave-triad interactions, leading to condensation of energy into long-wavelength trapped-ion structures. While numerous studies have independently reported regimes dominated either by strong zonal flows leading to reduced turbulence levels or by streamer-dominated states associated with elevated fluctuation amplitudes and enhanced radial transport, the transitional dynamics between these two states remain insufficiently  explored, particularly in controlled laboratory conditions. Thus, it is crucial to have the coexistence of both zonal flows and streamers, since their nonlinear competition governs the redistribution of turbulent energy and the resulting transport state of the plasma edge \cite{kasuya2008selective}. The balance between inverse cascade toward axisymmetric shear flows and anisotropic spectral transfer toward radially elongated structures determines whether turbulence is regulated or amplified. The simultaneous presence of ZFs and streamers implies a dynamically evolving energy landscape in $(k_r, k_\theta)$ space, where shear suppression and convective enhancement act concurrently. Understanding this coexistence is essential for clarifying the mechanisms underlying transport bifurcations, intermittency, and transitions between confinement regimes.
\noindent

In the present experiment, we report a controlled transition from a ZF-dominated regime to a streamer-dominated regime by systematically varying the ion-neutral collision frequency. Ion-neutral collisions modify the damping and drive balance of turbulence, thereby altering the nonlinear energy transfer pathways responsible for meso-scale structure formation. With increasing collisionality, we observe the coexistence of zonal flows and streamers, followed by a shift toward streamer like structure. This transition is followed by the modification of edge fluctuations, including changes in spectral distribution, intermittency signatures which might lead to convective transport. These results show that ion--neutral collisionality controls the reorganization of meso-scale structures and enables access to different self-organized turbulence states in edge magnetized plasma.\\
\noindent

The remainder of the paper is structured as follows. Section~II outlines the experimental configuration, probe diagnostics, and the data analysis techniques adopted in this work. Section~III details the observations in three representative regimes and its effect in edge fluctuations: (A) a zonal-flow dominated state at $R_m=32,~2\times10^{-5}\,\mathrm{mbar}$, (B) a regime where zonal flows and streamers coexist at $R_m=17,~5\times10^{-4}\,\mathrm{mbar}$, and (C) a streamer-dominated state at $R_m=51,~2\times10^{-3}\,\mathrm{mbar}$; it further presents (D) the characteristics of edge fluctuations and intermittency. Section~IV then interprets the main results and concludes with a summary.  
 
\begin{figure*}[]
\includegraphics[width=\linewidth]{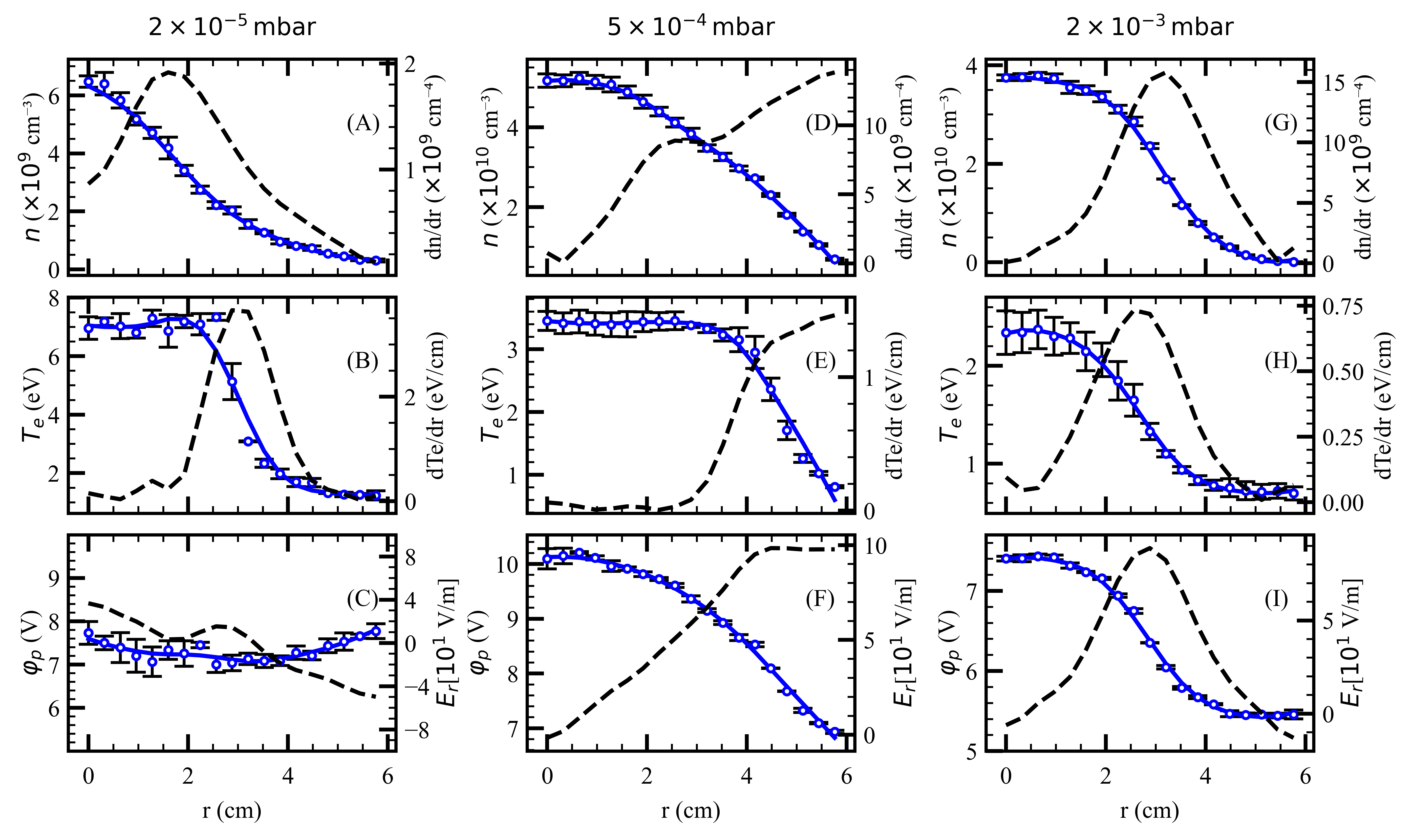} 
\caption{Radial mean profiles of electron density ($n$), electron temperature ($T_e$), and plasma potential ($\phi_p$) at different neutral pressures: (A--C) $2 \times 10^{-5}\,\mathrm{mbar}$, (D--F) $5 \times 10^{-4}\,\mathrm{mbar}$, and (G--I) $2 \times 10^{-3}\,\mathrm{mbar}$.
}     
\label{fig:Mean profile}  
\end{figure*}

\section{Experimental Setup, diagnostics and Data Acquisition }
\noindent

Experiments were carried out in the Inverse Mirror Plasma Experimental Device (IMPED), a linear magnetized plasma device that produces plasma using a two-dimensional array of ohmically heated tungsten filaments, as shown in Fig.~\ref{IMPED schematics}. The detailed description of the device and operating conditions has been reported previously in Refs.~\cite{bose_2015inverse,Bose_2015_RSI,roy2025_experimental,roy2025experimental}. The basic plasma parameters, namely ion density ($n_i$), electron temperature ($T_e$), and plasma potential ($\phi_p$), are measured using a single Langmuir probe (SLP) of 0.25~mm diameter and 5~mm length installed at port~F (Fig.~\ref{IMPED schematics}); the probe geometry is chosen to minimize gyro-orbit effects. The analysis procedure for obtaining $n_i$, $T_e$, and $\phi_p$ and their radial profiles follows the method described in our earlier work~\cite{roy2025_experimental,roy2025experimental}. In fluctuation measurements, ion saturation current fluctuations ($\tilde{I}_s$) are treated as density fluctuations ($\tilde{n}$), while temperature fluctuations ($\tilde{T}_e$), measured using a triple Langmuir probe, are found to be much smaller than floating-potential fluctuations ($\tilde{\phi}_f$) in IMPED, consistent with earlier observations in similar small-scale devices~\cite{roy2025_experimental,perks_2022impact,oldenburger_2012dynamics}. To investigate the spectral structure in $(k_r,k_\theta)$ space, several probe arrays (Fig.~\ref{Probe diagnostics}) are employed at different radial ports. At port~A, all four-tip array measures $\tilde{\phi}_f$, from which fluctuating $\mathbf{E}\times\mathbf{B}$ velocities are calculated: the radial velocity fluctuation $\tilde{v}_r=\tilde{E}_\theta/B_m$ is obtained from the poloidal electric field $\tilde{E}_\theta=-\partial\tilde{\phi}_f/\delta y$, and the poloidal velocity fluctuation $\tilde{v}_\theta=\tilde{E}_r/B_m$ from the radial electric field $\tilde{E}_r=-\partial\tilde{\phi}_f/\delta x$.  The spatial resolutions in the $x$ and $y$ directions are $\delta x = 4$~mm and $\delta y = 4.2$~mm for port A. The Reynolds stress $\langle \tilde{v}_r \tilde{v}_\theta \rangle$ is then evaluated to quantify nonlinear momentum transport and its role in zonal-flow generation. At port~B, another four-tip probe simultaneously measures $\tilde{n}$ and $\tilde{v}_r$ at the same virtual position and determines the radial wavenumber ($k_r$) from two radially separated ($\delta x=7~mm$) $\tilde{n}$ signals. At port~C, cross-phase analysis between density and potential fluctuations is performed; the poloidal wavenumber ($k_\theta$) is derived from the phase difference between signals measured at two poloidally separated tips ($\delta y=6.4~mm$), which helps identify mode structure and instability characteristics~\cite{smith_2007fast}. All probes are mounted on a digitally controlled drive capable of radial scanning with 0.2~mm positioning accuracy using a PID-based DC servo system~\cite{Roy2023_FED}. Fluctuating signals are digitized using a PXIe-based high-speed data acquisition system \cite{patel2025simultaneous} at 1~MS/s with 1~s records (1~M points), ensuring reliable fluctuation spectra and a low noise floor.
\begin{figure}[]
\includegraphics[width=\linewidth]{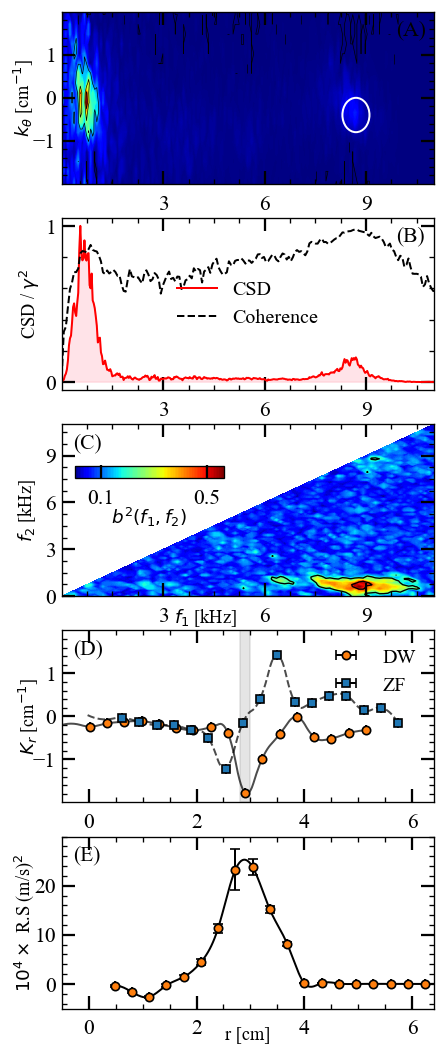} 
\caption{(A) $S(k,\omega)$ spectrum, (B) Cross power and squared coherence between radial velocity fluctuation ($\tilde{V_\theta}$) and floating potential fluctuation ($\tilde{\phi_f}$), (C) Auto-bicoherence analysis of $\tilde{\phi}_f$, (D) radial profile of $k_r$ in ZF (0.5-0.7 kHz) and DW frequency range ( 7-8 kHz), (E) Radial profile of turbulent Reynolds stress.}    
\label{fig:Identification of ZF at 2e-5 mbar}  
\end{figure}

\section{Experimental results}
\subsection{Zonal flow dominated regime}

Zonal flows (ZFs) are azimuthally symmetric, band-like shear flows that are ubiquitously observed in laboratory and fusion plasmas. Unlike mean shear flows driven by external momentum input or profile evolution, ZFs are self-generated by turbulence through Reynolds stress \cite{diamond_2004review,BDT_Theory_1990influence}. The Reynolds stress, defined as $\langle \tilde{v}_r \tilde{v}_\theta \rangle$, represents the correlation between fluctuating radial and poloidal velocities and acts as the driving force for the zonal flow \cite{Saikat_2018simultaneous}. Zonal flows are identified as low-frequency oscillations in the radial electric field with zero poloidal wavenumber ($k_\theta=0$) and finite radial wavenumber ($k_r$), accompanied by a radial polarity reversal. Their temporal evolution is governed by the balance between Reynolds stress drive and collisional damping, given by
\begin{equation}
\frac{\partial V_{\theta,\mathrm{ZF}}}{\partial t}
=
\frac{\partial}{\partial r}\langle \tilde{v}_\theta \tilde{v}_r \rangle
-
\gamma_{\mathrm{damp}} V_{\theta,\mathrm{ZF}},
\end{equation}
indicating that ZFs are preferentially destabilized in the low-collisionality regime \cite{diamond_2004review, tynan_2006observation}. Accordingly, experiments are performed at a neutral pressure of $2\times10^{-5}$~mbar and magnetic field of 550~G for $R_m=32$. This corresponds to a ratio of the ion--neutral collision frequency to the ion gyro frequency of $\nu_{in}/\Omega_{ci} = 0.02$. The radial profiles of $n$, $T_e$, and $\phi_p$ are shown in Fig.~\ref{fig:Mean profile}(A--C). The density gradient peaks at $r\simeq1.8$--3~cm, while the maximum $T_e$ gradient occurs around $r\simeq3$~cm. The plasma potential profile is relatively flat, ruling out the presence of strong mean shear flow and rulling out the possibility of Kelvin--Helmholtz-driven fluctuations. A low-frequency mode in the range of 600--700~Hz is observed along with the drift wave mode (m=2) of frequency 7-9 kHz (Fig \ref{fig:Identification of ZF at 2e-5 mbar}B). The drift mode is localized  in the density gradient region with comparable normalized $\tilde{n}/n$ and $\tilde{\phi}_f/kT_e$ \cite{jassby_1972transverse,schroder2005drift, roy2025_experimental}. The phase difference between $\tilde{n}$ and $\tilde{\phi}_f$ for this mode is less than $30^\circ$ \cite{brochard2005transition}. The drift mode is found to be nonlinearly coupled with the low frequency mode, as evidenced by the bicoherence analysis (Fig.~\ref{fig:Identification of ZF at 2e-5 mbar}C) \cite{kim_2007_digital}. This low-frequency mode which is dominant in the $\tilde{\phi_f}$ signal ($\frac{e\tilde{\phi}_f}{k_B T_e} > \frac{\tilde{n}}{n}$), modulates the drift-wave amplitude in time. The poloidal phase difference of two $\tilde{\phi}_f$ signals in that frequency range is nearly zero, confirming $k_\theta=0$, while the radial wavenumber, obtained from radially separated $\tilde{\phi}_f$ signals, reaches a maximum value of $\sim1.4~\mathrm{cm^{-1}}$ and exhibits a polarity reversal at $r\simeq2.8$~cm (Fig.~\ref{fig:Identification of ZF at 2e-5 mbar}D). Additionally, the radial wavenumber of the drift mode increases sharply near the ZF polarity reversal point, suggests tilting of large-scale drift eddies into smaller structures \cite{Terry_2000suppression, diamond_mean_flow_1991theory}. The low-frequency mode is spatially localized near $r=2$--3~cm, coinciding with a steep radial gradient of Reynolds stress (Fig.~\ref{fig:Identification of ZF at 2e-5 mbar}E). Those observations confirms that the low frequency mode is ZF, generated through nonlinear interactions among drift-wave modes. However, the dynamics between ZF and drift modes change with increasing neutral pressure. The modification in ion-neutral collisionality alters the non linear interaction coupling between these modes. This evolution of ZF and drift wave dynamics will be discussed in detail in the following section.  

\begin{figure}[]
\includegraphics[width=\linewidth]{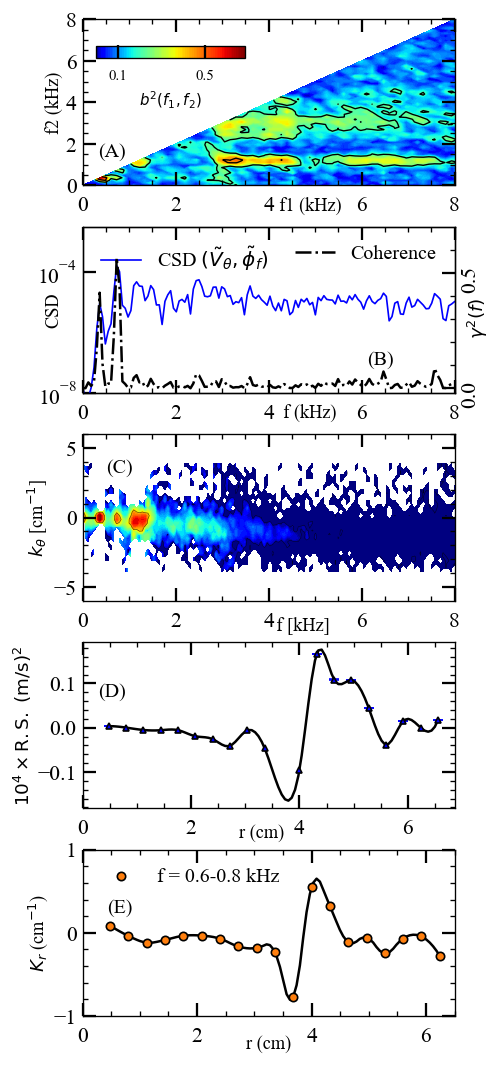} 
\caption{Identification of zonal flows (ZF) at $5 \times 10^{-4}$~mbar. (A) Auto bicoherence analysis of the floating potential fluctuation $\tilde{\phi}_f$. (B) Cross-power spectrum between the poloidal velocity fluctuation $\tilde{V}_\theta$ and $\tilde{\phi}_f$. (C) $S(k,\omega)$ spectrum. (D) Radial profile of the Reynolds stress. (E) Radial wavenumber ($k_r$) profile in the ZF frequency range.}    
\label{fig:Identification of ZF at 5e-4 mbar}  
\end{figure}

\subsection{Co-existence of zonal flow and streamer like structure}
\begin{figure}[]
\includegraphics[width=\linewidth]{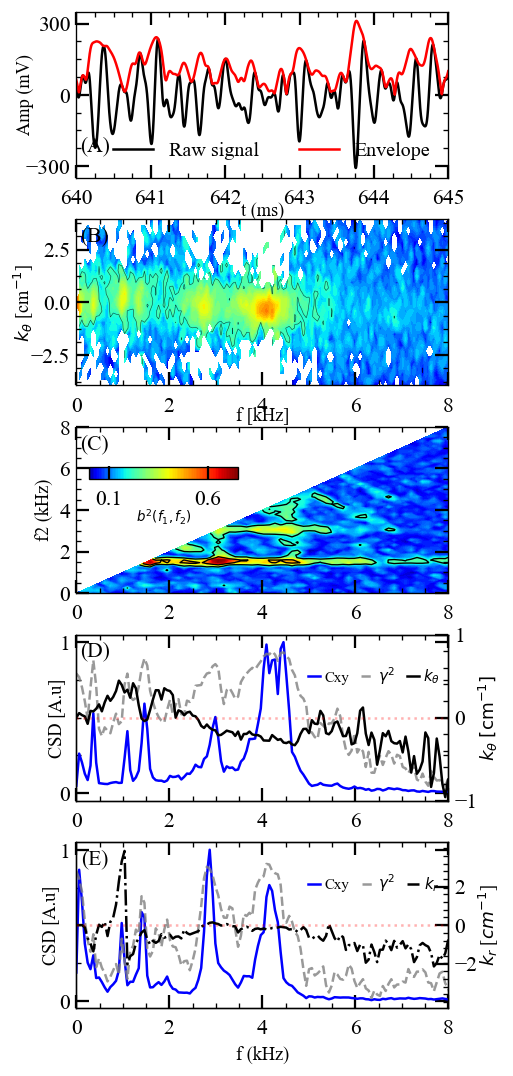} 
\caption{Identification of streamer at $5 \times 10^{-4}$~mbar. (A) Temporal evolution of the density fluctuation $\tilde{n}$ along with its envelope. (B) $S(k,\omega)$ spectrum. (C) Auto-bicoherence spectrum showing nonlinear coupling. (D) Power spectra of $\tilde{n}$ measured from two poloidally ($\theta$) separated probe signals together with the corresponding $k_\theta$ spectrum. (E) Power spectra of $\tilde{n}$ from two radially separated probe signals along with the corresponding $k_r$ spectrum.}   
\label{fig:Identification of streamer at 5e-4 mbar}  
\end{figure}

\paragraph{\textbf{Existence of ZF:}} Ion neutral collisions play an important role in damping ZF, as collisional friction reduces the sheared $E \times B$ flow energy of ZF \cite{diamond_2004review, itoh2006physics}. Therefore, it is expected that the ZF amplitude will decrease at higher neutral pressure. To examine the collisionality effect, the argon pressure is increased to $5\times10^{-4}~\mathrm{mbar}$. This corresponds to a ratio of the ion--neutral collision frequency to the ion gyro frequency of $\nu_{in}/\Omega_{ci} = 0.56$. A notable change in the magnitude and localization of the mean plasma profiles is observed (Fig.~\ref{fig:Mean profile}(D-F)). The peak density $n$ increases by nearly one order, and the position of its maximum gradient shifts to r = 4-5~cm. A similar trend is seen in the $T_e$ profile also. The peak plasma potential $\phi_p$ also increases from 7~V to 10~V, leading to the formation of a stronger radial electric field $E_r \approx 100~\mathrm{V/m}$ and a corresponding $E \times B$ velocity of approximately 1818~m/s. Spectral analysis is performed (Fig \ref{fig:Identification of ZF at 5e-4 mbar}, \ref{fig:Identification of streamer at 5e-4 mbar}) to identify the nature of the observed fluctuations. Fig~\ref{fig:Identification of ZF at 5e-4 mbar}A shows the presence of two low-frequency modes at 320~Hz and 700~Hz with coherence lebel 0.4 and 0.55 respectively, well above the significance level of $\approx0.01$. This two modes are localized at the r= 4.1~cm.  In this frequency range, the  $\tilde{\phi}_f/T_e$ is 5 times larger than the  $\tilde{n}/n$ . The poloidal velocity fluctuation $\tilde{v}_{\theta}$ exhibits finite coherence with $\tilde{\phi}_f$ oscillations at this frequency range (Fig.~\ref{fig:Identification of ZF at 5e-4 mbar}B), confirming their association with poloidal flow oscillations \cite{nagashima_2008coexistence}. The poloidal wavenumber $k_\theta$ is calculated using two poloidally separated $\tilde{\phi}_f$ signals and is shown in Fig.~\ref{fig:Identification of ZF at 5e-4 mbar}C. The value of $k_\theta$ is nearly zero in the given low-frequency range, satisfying the poloidal symmetry of the modes. Furthermore, Fig.~\ref{fig:Identification of ZF at 5e-4 mbar}(D,E) presents the radial profile of the Reynolds stress and the radial wavenumber $k_r$ in the given frequency range. A positive gradient of the Reynolds stress is observed around 4.1~cm, where the ratio $\tilde{\phi}_f/T_e$ to $\tilde{n}/n$ is maximized (5 times). At the same radial location, the radial wavenumber $k_r$ of the low-frequency mode changes its polarity. Thus, the simultaneous presence of $k_\theta \approx 0$, finite and polarity-reversing $k_r$, finite coherence with $\tilde{v}_{\theta}$, and a radial asymmetric Reynolds stress drive suggests that these low-frequency modes correspond to ZFs \cite{Terry_2000suppression, diamond_mean_flow_1991theory, BDT_Theory_1990influence, nagashima_2008coexistence}. The zonal flow (ZF) exhibits finite nonlinear interaction with the broadband turbulence spectrum, evident from the auto-bicoherence analysis (fig. \ref{fig:Identification of ZF at 5e-4 mbar}A). Although the interaction strength is relatively weak, it remains well above the noise floor ($\sim 0.01$). The mode amplitude is reduced compared to the $2\times10^{-5}$~mbar case. With increasing ion-neutral collision frequency, as the collisional damping of ZFs becomes stronger, which weakens their shear-regulating effect on turbulence. Thus, there is a possibility that the spectral energy may redistributed from $k_\theta = 0$ modes to $k_\theta \neq 0$, indicating an anisotropic transfer of energy that favors the formation of radially elongated structures \cite{fujisawa_2004identification, kasuya2008selective}. Under such conditions, streamer-like structures are expected to emerge, which may increase convective transport \cite{kasuya2008selective, kasuya_streamers_2010selective}. This will be discussed in the following paragraph. \\ 

 \paragraph{\textbf{Existence of streamers like structure:}}
The raw $\tilde{n}$ signal filtered in the 1 to 10~kHz frequency range, together with its envelope at the density gradient location ($r = 3.84$~cm) (Fig.~\ref{fig:Mean profile}D), is shown in  Fig.~\ref{fig:Identification of streamer at 5e-4 mbar}A. The signal exhibits quasi-periodic oscillations whose amplitude slowly modulates in time. The amplitude of high-frequency fluctuation modulates with 1-1.4 kHz, indicating amplitude modulation of higher-frequency components by a lower-frequency mode. Fig~\ref{fig:Identification of streamer at 5e-4 mbar}B shows the $k_\theta$ spectrum, where distinct modes are observed up to 5~kHz. To understand the spectral structure of the modes, the auto-power spectrum of $\tilde{n}$ is measured using two poloidally and radially separated probe tips from port C and port B, respectively. The corresponding $k_\theta$ and $k_r$ spectra is shown in Fig.~\ref{fig:Identification of streamer at 5e-4 mbar}(D-E). In the frequency range below 1~kHz, the mode has $k_\theta \approx 0$, which shows the ZF character in $\tilde{n}$ fluctuation also along with $\tilde{\phi}_f$. In contrast, modes in the 2 to 5~kHz frequency range exhibit finite $k_\theta$. A broad frequency band in the range 2-5 kHz along with two coherent peak of 3 and 4.1~kHz is observed, corresponding to $m=1,~m=2$ respectively. These modes rotating in the electron diamagnetic direction ($v_{De}$). Those modes have comparable fluctuation label ($\tilde{n}/n~ \sim~ \tilde{\phi}_f/T_e$).  Theoretically, the estimated Doppler shifted drift frequency \cite{schroder2005drift} is 4 to 8.1 kHz ($T_e=3~eV,~L_n~=2~cm,~E_{V\times B}=~80~V/m$ at $r=3.84~cm$), which is close to the frequency observed in the $\tilde{n}$ spectra. Thus, these modes are possibly linearly unstable drift waves \cite{horton1999drift,  schroder2005drift,roy2025_experimental,  karmakar2026zonal}. In addition to these drift modes, a lower-frequency mode of 1.1~kHz with $m=-1$ is also present, propagating in the ion diamagnetic direction, which is comparable to the envelope frequency of $\tilde{n}$ signal in Fig \ref{fig:Identification of streamer at 5e-4 mbar}A. This low-frequency mode is commonly referred to as the mediator \cite{kobayashi2017phenomenological, kasuya_streamers_2010selective, kasuya2008selective}, which is the key candidate for forming a streamer structure. The mediator is nonlinearly driven by drift waves and contributes to the broadening of the drift-wave spectrum. The interaction between the drift waves and the mediator leads to nonlinear broadening of the frequency spectrum, which is consistent with the observed amplitude modulation and nonlinear bunching in the spatiotemporal evolution of $\tilde{n}$. Such nonlinear bunching of drift waves results in the formation of convective cells \cite{yamada2010observation} which is radially elongated, as predicted in analytical studies \cite{nozaki1979solitons} of streamer formation. To quantify the nonlinear coupling, auto-bicoherence analysis is performed (Fig.~\ref{fig:Identification of streamer at 5e-4 mbar}C). A significant bicoherence value is observed between frequencies in the range of 1.5 to 5~kHz and 1 to 2~kHz, demonstrating nonlinear interaction between the drift waves and the mediator mode. The simultaneous presence of broadband drift-wave spectrum, a low-frequency mediator, nonlinear spectral broadening, and radially distributed finite non-linear coupling collectively establishes the identification of streamer-like structures \cite{kobayashi2017phenomenological, kin2019observations}.  Thus, in this pressure regime, zonal flows and streamer-like structures coexist at different radial locations in the plasma column. As ZF are subject to collisional damping~\cite{diamond_2004review}, an increase in ion neutral collisionality can further weaken ZF shear. The will be further examined in the following section.

\begin{figure}
\includegraphics[width=\linewidth]{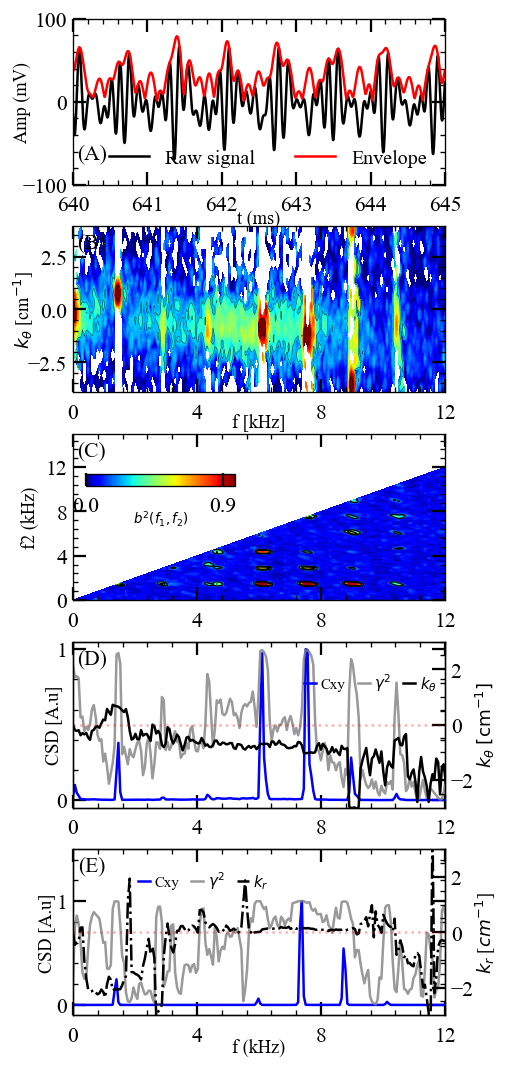} 
\caption{Identification of streamer at $2 \times 10^{-3}$~mbar. (A) Temporal evolution of the density fluctuation $\tilde{n}$ along with its envelope. (B) $S(k,\omega)$ spectrum. (C) Auto-bicoherence spectrum showing nonlinear coupling. (D) Power spectra of $\tilde{n}$ measured from two poloidally ($\theta$) separated probe signals together with the corresponding $k_\theta$ spectrum. (E) Power spectra of $\tilde{n}$ from two radially separated probe signals along with the corresponding $k_r$ spectrum.}    
\label{fig:Identification of streamer at 2e-3 mbar}  
\end{figure}

\begin{figure}
\includegraphics[width=\linewidth]{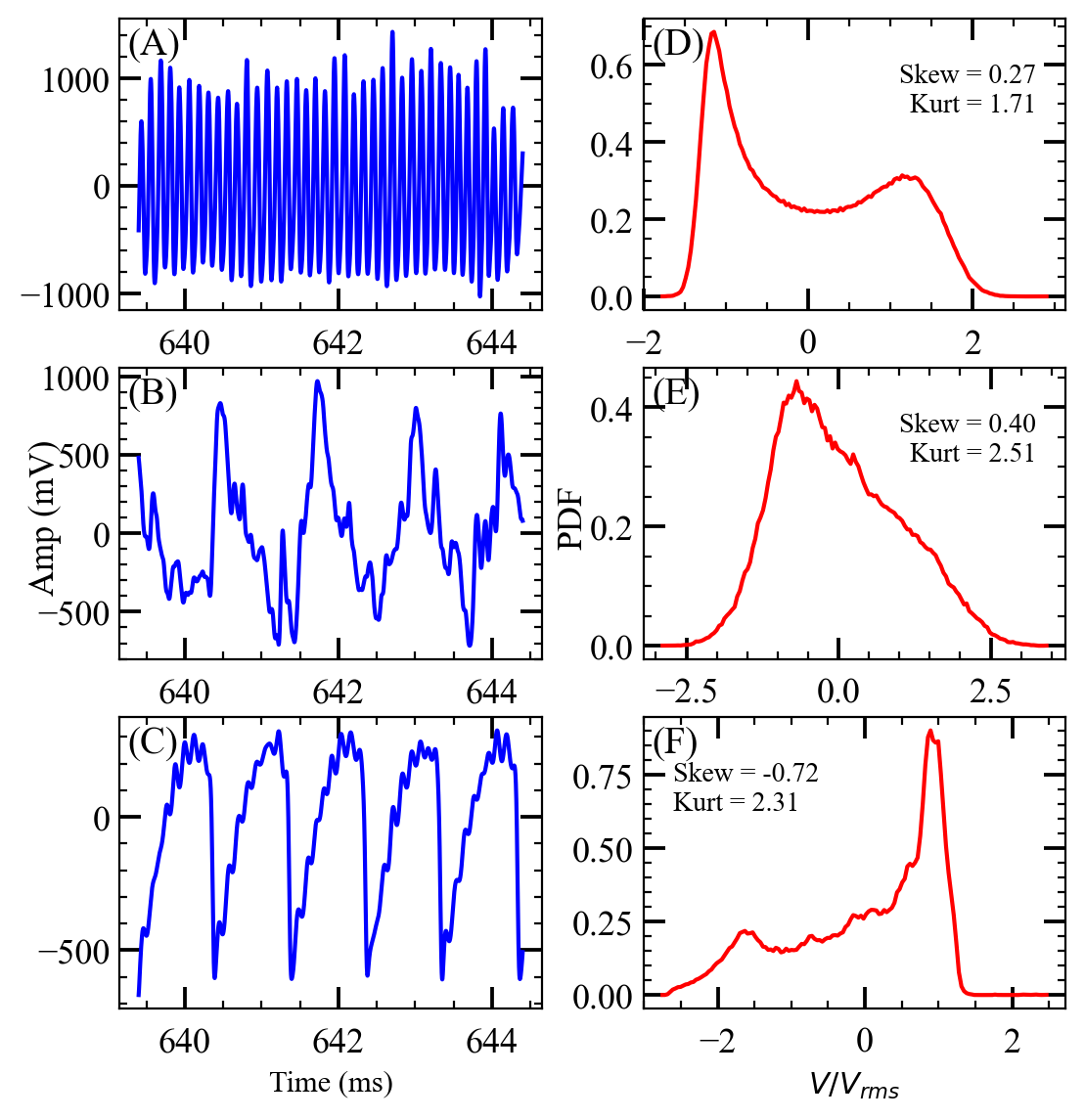} 
\caption{Raw $\tilde{n}$ signals (A--C) and their corresponding probability density functions (PDFs) (D--F), measured at $r = 4.5$~cm for different neutral pressures: (A, D) $2\times10^{-5}$~mbar, (B, E) $5\times10^{-4}$~mbar, and (C, F) $2\times10^{-3}$~mbar.}    
\label{fig:Edge fluctuation comparison}  
\end{figure}

\subsection{Streamer dominated regime}
 
\noindent
In this case, the ion neutral collision frequency is further increased by raising the argon pressure from $5\times10^{-4}~\mathrm{mbar}$ to $2\times10^{-3}~\mathrm{mbar}$. As a result, the ratio of the ion--neutral collision frequency to the ion gyro frequency is $\nu_{in}/\Omega_{ci} = 1.35$. This shifts gradient locations away from the edge (Fig \ref{fig:Mean profile}(G-I). From the fig \ref{fig:Mean profile}, it can be seen that, the peak  density does not vary monotonically with increasing neutral pressure, likely because both the plasma source conditions and the ion neutral collision frequency change simultaneously as the pressure is varied  \cite{kobayashi2017phenomenological}. With increasing pressure, density gradient increases. This increase in density gradient along with ion neutral collision substantially modifies the fluctuation characteristics of both $\tilde{n}$ and $\tilde{\phi}_f$. Across all radial locations, the fluctuation levels of $\tilde{n}$ and $\tilde{\phi}_f$ become comparable; however, downstream of the velocity shear location, a sudden enhancement in $\tilde{n}$ is observed. Within the radial range of 1 to 3.5~cm, dominant fluctuations extending up to 15~kHz are observed, indicating broadband turbulent fluctuation. The raw $\tilde{n}$ signal along with its envelope is shown in  Fig.~\ref{fig:Identification of streamer at 2e-3 mbar}A. Substantial amplitude modulation, which is more prominent than the case of $5 \times 10^{-4}~mbar$ is visible. This envelope is quasi-periodic in nature, which shows the presence of the streamer as bunching of drift waves \cite{kobayashi2017phenomenological, kin2019observations, yamada_2008anatomy}. Spectral analysis (Fig.~\ref{fig:Identification of streamer at 2e-3 mbar}D) shows that the high frequency modes in the frequency range 3.5-9 kHz propagate in the electron diamagnetic direction ($v_{De}$). The amplitude of these drift modes is modulated at approximately 1.3~kHz, which closely matches the envelope frequency of about 1.45~kHz observed in the time trace (Fig.~\ref{fig:Identification of streamer at 2e-3 mbar}A), indicating modulation by a low-frequency mediator.  This mediator mode is rotating in ion diamagnetic direction. Multiple coherent modes are present up to 11~kHz (Fig.~\ref{fig:Identification of streamer at 2e-3 mbar}B). The mediator frequency ($\sim$1.45~kHz) nonlinearly interacts with different drift modes (6.12, 7.5, 8.9, and 10.5~kHz), as confirmed by the auto-bicoherence spectrum (Fig.~\ref{fig:Identification of streamer at 2e-3 mbar}C), which demonstrates multiple non-linear triad interactions between the mediator and higher-frequency drift waves.
Such nonlinear triad interactions between neighboring drift modes mediated by a low-frequency mode have been theoretically shown to generate radially elongated convective cells (streamers) \cite{kasuya_streamers_2010selective, kasuya2008selective}. Through this nonlinear process, the bunched structure is sustained for a duration much longer than the characteristic drift-wave oscillation period. The radial wavenumber spectrum $k_r$ (Fig.~\ref{fig:Identification of streamer at 2e-3 mbar}E) further shows that modes in the 3.5 to  9~kHz have $k_r \approx 0$, satisfying the streamer condition of radially elongated structure. Overall, the key spectral features are observed in the present measurements: (i) a typical scale with finite poloidal wavenumber ($k_\theta$) and nearly zero radial wavenumber ($k_r \sim 0$), (ii) nonlinear coupling of adjacent drift modes through the mediator mode forming a quasi-mode like structure, and (iii) drift modes and the mediator mode propagates in the opposite direction (drift modes propagates in electron diamagnetic direction and the mediator mode moves in ion diamagnetic direction). These are consistent with presence of streamer like structure. The identification of streamer-like  structures in the present work is based on a combination of experimentally established spatial and temporal  characteristics \cite{kobayashi2017phenomenological}, rather than solely on direct measurements of long-range spatial correlation. Direct estimation of large-scale correlations is not feasible with two or three tip probes. Instead, streamer-like dynamics are inferred from signatures consistent with $k_r \sim 0$ and finite $k_\theta$, indicating radially elongated and poloidally structured fluctuations \cite{kasuya2008selective, kasuya_streamers_2010selective, diamond_2004review}. Additionally, the presence of quasi-periodic temporal oscillations and broadband spectral features suggests strong nonlinear dynamics. In particular, the non-linear interaction between adjacent drift-wave modes propagating in electron diamagnetic direction mediated by a lower-frequency mode propagating in the opposite propagation direction is observed, which is essential for the formation of streamer structures and is identified as the mediator mode \cite{yamada2010observation}. Such nonlinear interactions and the associated spectral broadening have been identified in previous  experimental \cite{yamada_2008anatomy, kobayashi2017phenomenological} and theoretical studies \cite{kasuya2008selective} as key features linked to streamer formation. Although direct measurement of long-range spatial correlation is beyond the scope of the present diagnostic configuration, the observed combination of these signatures provides indirect but consistent evidence for streamer-like dynamics. As the ion neutral collision frequency normalized to the ion gyrofrequency increases from 0.02 to 0.56 and further to 1.56, the turbulence regime evolves from zonal-flow (ZF) dominated, to a coexistence of ZF and streamer structures, and finally to a fully streamer-dominated state \cite{kasuya2008selective}. With a substantial increase in ion-neutral collisionality, a transition from zonal flows (ZFs) to streamer-like structures is observed. Such a change in the nonlinear state is expected to influence edge fluctuations. This aspect is discussed in the following section.

\subsection{Edge Fluctuation and Intermittency}
\noindent
Fig \ref{fig:Edge fluctuation comparison} shows the time evolution of edge density fluctuations ($\tilde{n}$) at three different neutral pressures, increasing from $2\times10^{-5}$~mbar, $5\times10^{-4}$~mbar, and $2\times10^{-3}$~mbar. The edge region is defined as the radial location where the plasma density decreases to approximately $1/e$ of its peak value. Based on this, the edge is located at $\sim r = 5.44$ cm at $2\times 10^{-5}$ mbar and $5\times 10^{-4}$ mbar, while for $2\times 10^{-3}$ mbar, it shifts inward to $\sim r = 4.5$ cm. As discussed in the previous section, increasing the ion neutral collision frequency drives the system from a zonal-flow (ZF) dominated regime to a coexistence regime (ZF + streamer), and eventually to a streamer-dominated state. At the lowest pressure ($2\times10^{-5}$~mbar), the signal exhibits relatively regular oscillations with weak amplitude modulation. The waveform appears more coherent and narrowly distributed in frequency, overall density fluctuation label is high enough ($\approx40\%$). At the intermediate pressure ($5\times10^{-4}$~mbar), signal is rightly skewed and consists of low frequency coherent oscillation. However, density fluctuation is reduced to $5-6\%$ \cite{kasuya2008selective}. Increasing pressure further results ($2\times10^{-3}$~mbar)  sawtooth-like fluctuation characterized by a slow ramp-up phase followed by a rapid ramp-down phase. Overall density fluctuation maintains to $6\%$. The pattern of the waveform suggests possible presence of a nonlinear wave \cite{liu_2018experimental}. However, a detailed identification and characterization of this wave are left for future investigation. The turbulent de-correlation time is a measure to quantify the temporal coherence and lifetime of fluctuating structures in the plasma. It represents the characteristic time over which fluctuations remain correlated and thus provides a measure of how long turbulent eddies or coherent structures persist. In drift-wave-dominated turbulence, the de-correlation time is typically short due to rapid phase mixing and wave propagation, whereas the emergence of long-lived coherent structures is associated with a significant increase in this time scale. The turbulent de-correlation time is calculated for the corresponding edge fluctuations.  It is observed that with increasing neutral pressure, the turbulent de-correlation time increases significantly, from $24~\mu$s to $30~\mu$s and further to $497~\mu$s. At the highest pressure, the auto-decorrelation time becomes much larger than the characteristic drift-wave time scale, indicating a substantial modification in the temporal dynamics of the turbulence and suggesting the presence of long-lived coherent structures at higher ion-neutral collisionality. However, with increasing ion-neutral collision frequency, the background plasma profiles are also modified,  which act as the primary drive for the fluctuations. Therefore, the observed edge transition in temporal dynamics cannot be attributed solely to collisional effects; rather, it arises from the combined influence of increased collisionality and the concurrent modification of background gradients, both of which play an important role in shaping the turbulence timescale. Such temporally long-lived structures are consistent with streamer-like dynamics which might facilitate enhanced convective radial transport. Fig \ref{fig:Edge fluctuation comparison}(D-F) shows the probability distribution functions (PDFs) of the edge density fluctuations corresponding to increasing neutral pressure. At the lowest pressure ($2\times10^{-5}$~mbar), where the plasma is in a ZF-dominated regime, the PDF is  weakly skewed (Skew = 0.27, Kurt = 1.71), indicating more symmetric and less intermittent fluctuations. This is consistent with the raw time trace, which shows more regular oscillations due to the shear regulation by zonal flows. At the intermediate pressure ($5\times10^{-4}$~mbar), corresponding to the coexistence of ZF and streamer structures, the PDF becomes broader and more positively skewed (Skew = 0.40, Kurt = 2.51). The increase in skewness and kurtosis reflects enhanced intermittency and burst-like events, which are also visible in the raw signal as amplitude modulation and envelope formation. Increasing pressure further results ($2\times10^{-3}$~mbar) streamer dynamics to dominate, the PDF becomes strongly asymmetric (Skew = -0.72, Kurt = 2.31) with a pronounced negative tail, indicating large-amplitude intermittent events at the edge. Thus, the evolution of the edge density fluctuation reflects the underlying transition of meso-scale dynamics: from shear-regulated turbulence (ZF), to mixed dynamics (ZF + streamer), and finally to streamer-dominated convective transport as neutral pressure increases.   
 
\section{Discussion and Conclusions}

\noindent
In the present experiment, the ion--neutral collision frequency is used as a control parameter to vary the plasma collisionality. At low pressure ($2\times10^{-5}~\mathrm{mbar}$), a zonal flow (ZF) in the range of 600--700~Hz is nonlinearly excited by drift-wave turbulence. As the pressure increases to $5\times10^{-4}~\mathrm{mbar}$, the growth of the ZF is reduced \cite{diamond_2004review,kim2002_dynamics}, while the growth of streamers is enhanced through nonlinear interactions between neighboring drift modes via a mediator mode. A further increase in pressure to $2\times10^{-3}~\mathrm{mbar}$ leads to substantial damping of the ZF and the prominent development of streamers via multiple nonlinear interactions among drift modes along with the mediator mode. Previous theoretical studies have shown that streamer-like structures can drive significant radial transport in the SOL region. This understanding is further supported by experimental observations in JT-60U \cite{kin2023experimental}, DIII-D \cite{hong2023observation}, KSTAR \cite{choi2019experimental}, and HL-2A \cite{cheng2020formation}. With increasing pressure, the dynamics of edge fluctuations also change significantly. In the low-pressure regime, fluctuations are dominated by regular oscillations, and the corresponding probability density function (PDF) is weakly intermittent. In contrast, with increasing ion--neutral collision frequency, the total fluctuation energy in $\tilde{n}$ and $\tilde{\phi}_f$ decreases, while the fluctuations become more asymmetric and intermittent. The characteristic scale of the fluctuations increases, and the fluctuation energy shifts toward low-frequency oscillations responsible for convective transport \cite{Xu_10.1063/1.3302535}. Turbulence dominated either by zonal flows or by streamers is observed in the nonlinear saturation stage. Nonlinear coupling between modes with neighboring azimuthal mode numbers $m$ and close frequencies leads to streamer formation, and the streamer observed in this experiment exhibits a timescale different from that of the drift wave. These results demonstrate the selective formation of zonal flows and streamers using the ion--neutral collision frequency as a control parameter. The collision frequency represents the effective damping strength of turbulent structures. When the collision frequency is large, zonal flows are strongly damped and streamer structures dominate. At the edge, although the fluctuation amplitude is relatively small in this regime, the fluctuations occur at larger spatial scales. Conversely, when the ion--neutral collision frequency is small, the fluctuation amplitude is large but the fluctuations are less intermittent, likely due to the shearing action of zonal flows on drift-wave turbulence. Thus, the characteristics of edge fluctuations are strongly governed by the underlying turbulent structures.

\begin{acknowledgments}
The authors would like to express their sincere gratitudes to Kalpesh Doshi, Jignesh Patel, Kirti Mahajan, Manisha Bhandankar, Ritesh Sugandhi, Minsha Shah, Praveenlal, and Sandip Das from the Institute for Plasma Research (IPR), India, for their valuable experimental support. AS is grateful to the Indian National Science Academy (INSA) for the INSA Honorary Scientist position.
 
\end{acknowledgments}

\bibliography{References}

\end{document}